\documentclass[final, runningheads]{llncs}
\usepackage[T1]{fontenc}
\usepackage{graphicx}
\usepackage{subcaption}
\usepackage{hyperref}
\usepackage{amsmath}
\newcommand{\itadata}{\footnotesize \textsl{ITADATA2024: The 3$^{\text{rd}}$ Italian Conference on Big Data and Data Science}}
\usepackage{fancyhdr}
\fancypagestyle{empty}{%
  \fancyhf{}
  \fancyhead[C]{\itadata}
  \fancyhead[R]{\includegraphics[width=.05\textwidth]{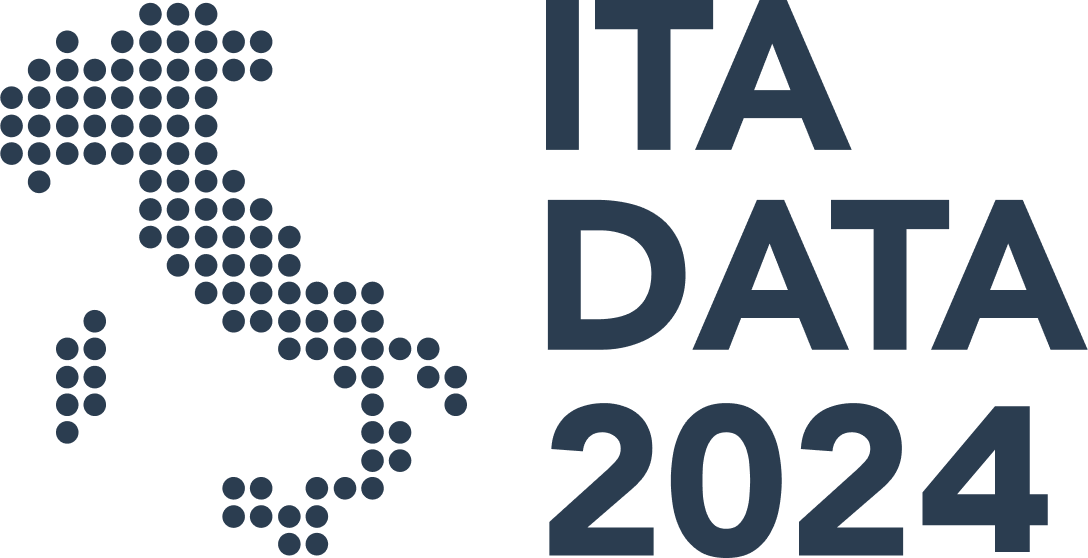}}
}

\usepackage{xcolor}
\usepackage{cleveref}
\usepackage{booktabs}

\def\ie{i.e.,~}               

\begin{document}
\title{Exploring Few-Shot Object Detection on Blood Smear Images: A Case Study of Leukocytes and Schistocytes}
%
%
\author{Davide Antonio Mura, Michela Pinna, Lorenzo Putzu, \\Andrea Loddo, Alessandra Perniciano, Olga Mulas, Cecilia Di Ruberto}
\authorrunning{Mura et al.}
%
\institute{University of Cagliari, Cagliari, Italy\\
\email{m.pinna116@studenti.unica.it, \{davideantonio.mura,lorenzo.putzu,andrea.loddo,\\alessandra.pernician,olga.mulas,cecilia.dir\}@unica.it}}
\maketitle              
\begin{abstract}
The detection of blood disorders often hinges upon the quantification of specific blood cell types. Variations in cell counts may indicate the presence of pathological conditions. Thus, the significance of developing precise automatic systems for blood cell enumeration is underscored.

The investigation focuses on a novel approach termed DE-ViT. This methodology is employed in a Few-Shot paradigm, wherein training relies on a limited number of images. Two distinct datasets are utilised for experimental purposes: the Raabin-WBC dataset for Leukocyte detection and a local dataset for Schistocyte identification. In addition to the DE-ViT model, two baseline models, Faster R-CNN 50 and Faster R-CNN X 101, are employed, with their outcomes being compared against those of the proposed model.

While DE-ViT has demonstrated state-of-the-art performance on the COCO and LVIS datasets, both baseline models surpassed its performance on the Raabin-WBC dataset. Moreover, only Faster R-CNN X 101 yielded satisfactory results on the SC-IDB. The observed disparities in performance may possibly be attributed to domain shift phenomena.

\end{abstract}
\section{Introduction}
\label{sec:intro}
Haematology is the medical field that deals with the diagnosis and treatment of blood-related diseases which may affect the quantity and functionality of blood cells, blood coagulation, or even the immune system. Within the bloodstream are three main cell types: platelets or thrombocytes, red blood cells (RBCs) or erythrocytes, and white blood cells (WBCs) or leukocytes. An increase or decrease in the amount of these cells can lead to haematological diseases, such as anaemia caused by a reduction in RBCs or leukocytosis caused by an increase in WBCs. Haematologists' skill in accurately counting specific blood cell types is paramount. However, since humans are inherently prone to error, developing automatic or semi-automatic systems for cell counting is essential.

Other works in the literature propose methods for blood cell counting, including a deep convolutional generative adversarial network (DCGAN) for classifying and detecting WBC~\cite{ref_related_1} or several deep learning (DL) models, such as VGG-16, Efficient Net, ResNet-50, Inception, and Xception, tested for WBC classification~\cite{ref_related_2}. In these works, however, datasets containing numerous images were used, which made it possible to exploit transfer learning to classify images correctly.

Based on these premises, the motivation behind this work is that the available labelled data is often limited in the medical field, especially in medical imaging. Labelling a large amount of data, such as that required today to train DL algorithms, requires considerable time and effort. Also, it must be considered that different datasets typically have various characteristics due to the different acquisition conditions or protocols, causing the domain shift problem. In this context, Few-Shot methods can contribute to solving the problems described~\cite{ref_fewshot1,ref_fewshot2}.  

This study presents various methodologies for identifying blood cells within microscope images. We will delve into Object Detection (OD), a technique employed to detect objects such as cells, explicitly focusing on Few-Shot Object Detection (FSOD). The analysis will examine and test an FSOD method known as DE-ViT~\cite{ref_devit}. We compared its effectiveness with traditional OD techniques to provide a classic reference baseline and evaluate the robustness and reliability of established methods in scenarios where labelled data are scarce.

This work focused on two main tasks: 1) the detection of the five main WBC sub-types, with potential implications in leukaemia or WBC-related diseases identification~\cite{ref_ai2030025}, 2) the detection of schistocytes, a particular type of RBC associated with diseases such as thrombotic thrombocytopenic purpura (PTT). PTT is a severe disease involving micro-blood clots throughout the body, obstructing blood flow to vital organs such as the brain, heart and kidneys~\cite{ref_schisto}.

Both tasks were tackled by considering a situation in which insufficient data was available to use classical DL models. In task 1), the dataset used contains numerous images, but to comply with the constraint imposed as a premise, a subset of the dataset was selected. Whereas for task 2), being a new dataset, no other annotated public dataset was found, so the constraint imposed was respected a priori. Moreover, this choice is also motivated by the fact that we wish to create a system adaptable to any dataset and which can address the problem of domain shift, which plagues medical images due to the type of acquisition, which can vary for each hospital~\cite{ref_related_3,ref_related_4}. 

\section{State of the Art}
\subsection{Background}
As mentioned, blood primarily comprises WBCs, RBCs, and platelets. WBCs are immune system cells that defend the body against microorganisms and foreign bodies. They feature a nucleus enveloped by cytoplasm, which houses various cell organelles. Conversely, erythrocytes, or RBCs, are anucleate and exhibit a biconcave disc shape optimized for oxygen transport to tissues. Platelets, or thrombocytes, are small, nucleus-free cell fragments that play a critical role in clot formation following vascular injury.

Under normal conditions, mature blood cells are released into the bloodstream, primed for their respective functions. However, various pathological conditions can alter the number and type of blood cells, releasing immature or malformed cells. These aberrant cells may be functionally deficient or entirely non-functional. Consequently, blood cell counting is a crucial diagnostic practice, enabling the comparison of cell counts in a blood sample against established reference ranges to ascertain whether the proportions of different cell types are within normal limits.

\begin{figure}[h]
\centering
\includegraphics[scale=0.12]{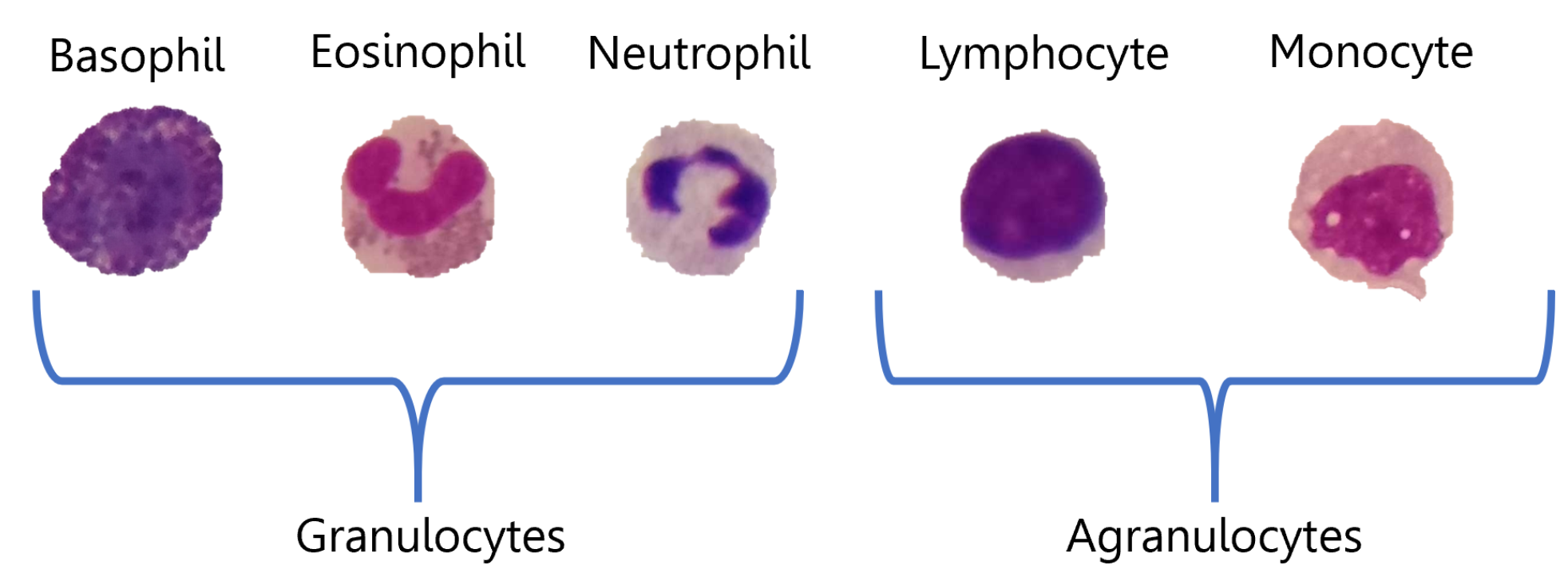}
\caption{Image depicting the 5 types of leucocytes~\cite{ref_ai2030025}.}
\label{fig:wbc_type}
\end{figure}

A significant challenge in studying WBCs lies in their morphological diversity, complicating their analysis and identification. As illustrated in~\Cref{fig:wbc_type}, WBCs can be categorised into two primary groups: granulocytes and agranulocytes. Granulocytes, characterised by granules, include neutrophils, eosinophils, and basophils. Agranulocytes, which lack granules, consist of monocytes and lymphocytes. Differentiation among these WBC types is based on the presence or absence of granules and variations in size, shape, and the number of nuclear lobes.

Regarding RBCs, one problem related to them is Poikilocytosis, a condition in which numerous erythrocytes are present that do not have the usual shape and size. The quantity of these abnormal cells may indicate the presence of specific diseases or disorders. Schistocytes are one of the various forms of Poikilocytosis, which, unlike normal erythrocytes, have an angular shape, typically with two pointed ends. As a result of damage to a blood vessel, a clot begins to form, and fibrin filaments are produced. Consequently, RBCs become trapped in these filaments. The sharp force of the blood flow then tends to break up the trapped RBCs, resulting in fragments known as schistocytes.

\begin{figure}
\centering
\begin{subfigure}{.46\textwidth}
  \centering
  \includegraphics[width=\textwidth]{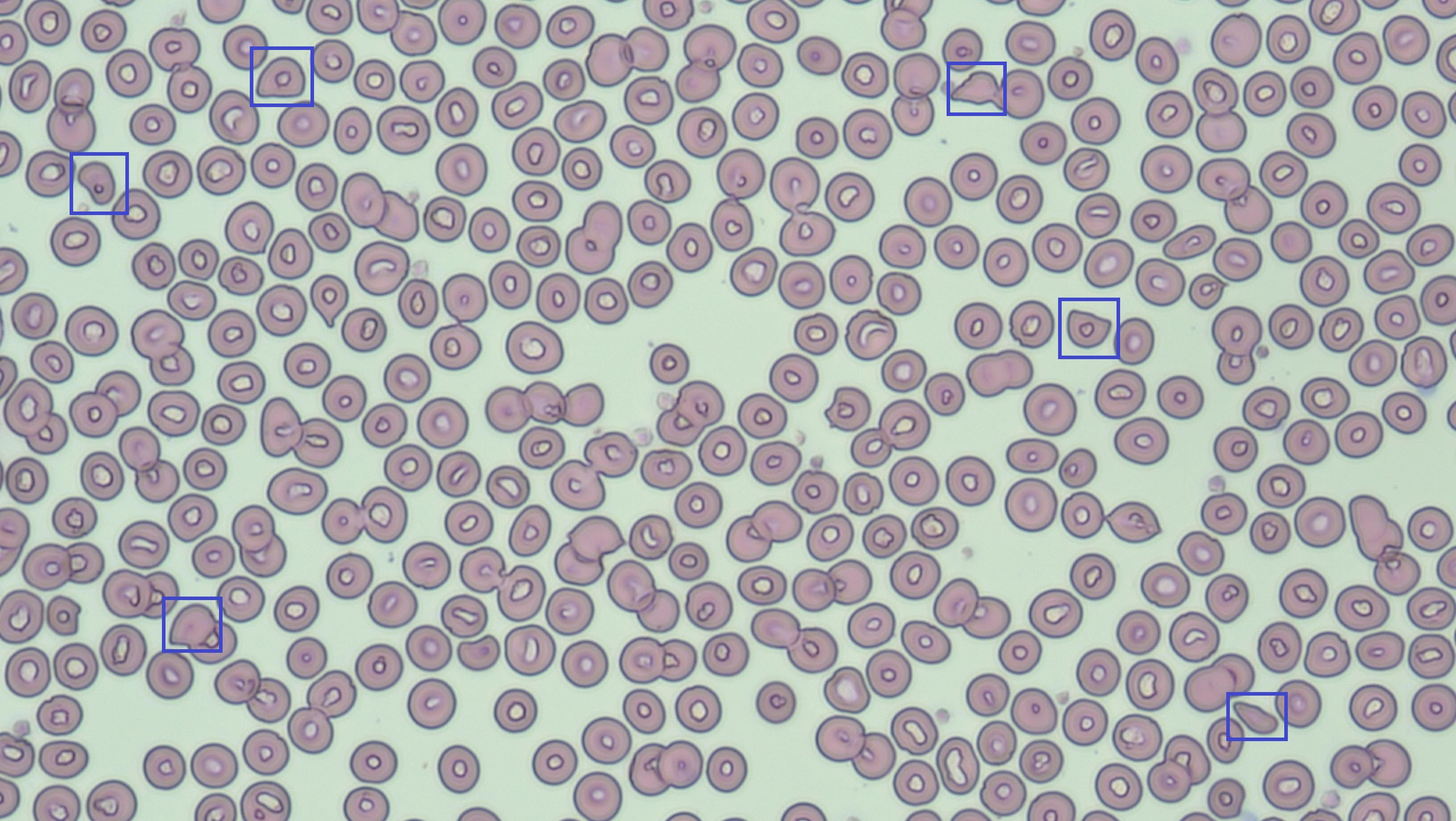}
  \caption{Full-size blood smear.}
  \label{fig:sub1}
\end{subfigure}%
\hfill
\begin{subfigure}{.475\textwidth}
  \centering
  \includegraphics[width=\textwidth]{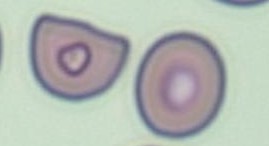}
  \caption{Schistocyte (left) and RBC (right).}
  \label{fig:sub2}
\end{subfigure}
\caption{Sample images extracted from the dataset involved in this work. \Cref{fig:sub1} presents a full-size blood smear image with some schistocytes inside the blue boxes. \Cref{fig:sub2} is a crop of the same image presenting a schistocyte and a healthy RBC. Source: SC Ematologia e CTMO Ospedale Businco Cagliari.}
\label{fig:schisto_sota}
\end{figure}

Detecting schistocytes and their count is important to diagnose possible diseases such as microangiopathic haemolytic anaemia, which is detected by a greater than 1\% presence of schistocytes in the smear. \Cref{fig:schisto_sota} shows a blood smear in which mainly erythrocytes are present, and bounding boxes highlight some schistocytes.

\subsection{Related Work}
The literature describes several approaches to blood cell detection and counting, mainly those based on segmentation and object detection.
Segmentation involves isolating one or more groups of pixels from the image, such that these groups are the objects to be identified within the image~\cite{ref_ai2030025}. 

Object detectors could be grouped into two main categories: two-stage detectors, called proposal-based, in which candidate regions are first extracted, which are potentially the objects to be identified and then they are improved and finally classified; one-stage detectors, called proposal-free, which locate and classify without first generating proposals. In general, the former have better results, in terms of accuracy of predictions, however, the latter turn out to be faster in the inference phase.

Focusing on two-stage detector methods, they present this architecture: from the input images, a feature map is first extracted, on which the detector identifies an object in its subregion, called Regions of Interest (ROIs); for each subregion, both the regression task, to find the bounding box that best encloses the object, and the classification task, which assigns a label to the object, are performed.

As is the case with numerous Computer Vision (CV) tasks, the latest Object Detection (OD) methods are predominantly based on Deep Learning (DL) models, which are known for their need for substantial amounts of training data. However, obtaining sufficient data is not always feasible. For instance, recognising rare diseases or identifying individuals via video surveillance often suffers from data scarcity. In such scenarios, few-shot learning (FSL) seeks to develop models capable of learning from only a few examples. An image classification task within the FSL framework, known as few-shot classification (FSC), involves a dataset with $N$ classes and $K$ images per class. This setup is also termed $N$-way-$K$-shot classification. A notable particular case of $K$-shot is one-shot classification, where there is only one image per class, or even zero-shot classification, where no examples exist for that class.

While FSC focuses on categorising an entire image with $K$ instances per class, FSOD adds the complexity of localising one or more objects within the image. This additional complexity means that FSC methods cannot directly apply to FSOD tasks. In a $K$-shot object detection task, the model aims to learn to classify and localise objects that belong to the respective classes with $K$ instances for each class.

The challenge of identifying different types of WBC using few-shot methods has been explored in a single work~\cite{ref_fewshot_wbc}, which introduced a classification method for Acute Myeloid Leukemia. This study proposed a few-shot learning framework consisting of a base classifier, a ResNet-18, and a meta-learning module to optimise performance. The results were promising, paving the way for experimenting with different few-shot approaches, such as the one proposed in this work.

However, to the best of our knowledge, no studies have addressed the identification of schistocytes using few-shot learning methods.

\section{Materials and Methods}

\subsection{Dataset}
Two datasets were used in this work: the Raabin-WBC~\cite{Kouzehkanan2021.05.02.442287} for WBC analysis and a novel dataset related to schistocytes. It will be named SC-IDB (SChistocytes Image Data Base).

\subsection{Raabin-WBC}
Raabin-WBC contains all five main types of WBCs: Lymphocytes, Monocytes, Neutrophils, Eosinophils, and Basophils. It also contains two further categories, namely Bursts and Artefacts, representing the cell burst during the blood smear analysis and the errors committed, such as dye residues in the slide. 

The dataset contains about 40,000 images, with a resolution of $2,988\times5,312$ pixels. Nevertheless, for the experiments, we worked in a Few-Shot configuration, so 97 images were used in total: 22 Monocytes, 27 Burst, 19 Small Lymphocytes, 18 Large Lymphocytes, 29 Neutrophils, 18 Eosinophils and 21 Artefacts. 

The images were systematically divided into three subsets while maintaining class balance: 34 images for the training set, 10 for the validation set, and 53 for the testing set. This subdivision ensures that tests are conducted on a specific subset of images, adhering to the initial constraints imposed in~\Cref{{sec:intro}}.

\subsubsection{Preprocessing.}
Before training, every image underwent a crop strategy since the Raabin-WBC images have a black frame, with the blood sample image in the centre. As the black frame is devoid of information, what has been done is to remove the black border by identifying the largest rectangle inscribed in a circle so that the resulting images are free of black borders. In addition, the images were resized to reduce their resolution to $794\times746$ pixels; this reduced the problem's dimensionality and accelerated the experiments' time. Finally, the cells were manually segmented to provide the method described below with greater localisation accuracy.

\subsection{SC-IDB}
The SC-IDB images present a particular type of RBC in addition to the classic WBCs and RBCs: the schistocyte. There are 124 images with a resolution of 3264x1840 pixels, and three experts labelled 260 schistocytes. The images were collected at the S.C. Haematology and CTMO of the Businco Hospital in Cagliari.
The dataset's split into training, validation, and test sets contained 35, 8, and 81 images, respectively.

\subsubsection{Preprocessing.}
In contrast to the previous dataset, the schistocyte images were not cropped as they were already free of black borders. However, due to the high resolution, a resize was performed that reduced the resolution of the images to approximately $979\times552$ pixels. All cells in the dataset were then manually segmented.

\subsection{Evaluation Approach}
In this work, an initial objective was to test the DE-ViT~\cite{ref_devit} model, an Open-Set Object Detector that uses image prototypes to categorise objects. The basic idea of the method is to extract a prototype for each class (category), \ie a generic class representation. In this way, features are extracted from the new input image during the inference phase and compared with the prototypes of the various classes. A pre-trained backbone DINOv2 ViT~\cite{oquab2023dinov2,darcet2023vitneedreg} was used for feature extraction and prototype creation.

DINOv2 is a self-supervised method, \ie it does not need labelled data.
Since the method is self-supervised, it can extract important information about the image context without labels. This will be seen to be a key aspect of the DE-ViT model. In turn, DINOv2 is based on Vision Transformer (ViT)~\cite{dosovitskiy2021image}, a model for image classification based on the Transformer architecture~\cite{vaswani2023attention}.

The method's strength is adding new classes without any training/fine-tuning procedure. To add new classes, prototypes must be extracted, a procedure unrelated to the model training.

\subsubsection{Prototype creation}
A fundamental element of the DE-ViT model architecture is using prototypes, \ie feature vectors representing a class. 

\begin{figure}[h]
\centering
\includegraphics[width=10cm, height=4cm]{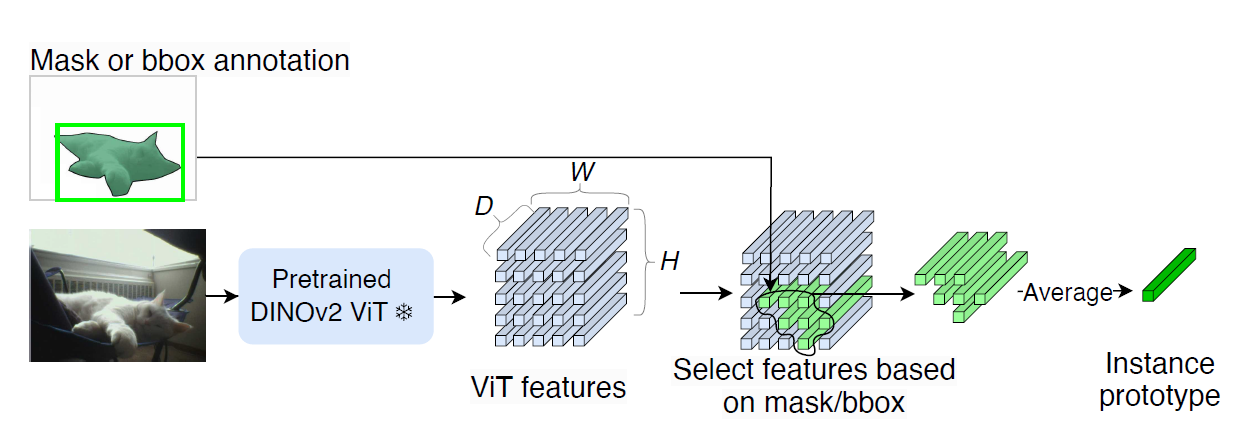}
\caption{Prototype creation scheme of the DE-ViT model. Source: Zhang et al.~\cite{ref_devit}}
\label{fig:devi4}
\end{figure}

Using the DINOv2-ViT backbone, image features are extracted from the support images. Only features corresponding to the bounding box or segmentation are averaged. Afterwards, clustering between all the prototypes for each extracted class is carried out, resulting in a single vector corresponding to the prototype and representing the class to be detected. \Cref{fig:devi4} shows the process diagram for prototype extraction. 

Background prototypes are also extracted similarly to class prototypes; however, the only difference is that background prototypes do not change but are fixed, becoming a parameter of the model. This decision is made by observing that backgrounds typically share similar characteristics, such as colours, shapes, etc.

It is also important to emphasise correctly separating foreground and background classes. For this reason, masks were used to extract the background prototypes to isolate the foreground elements from those in the background. The creation of the prototypes is an offline procedure compared to the training of the model.

\section{Experiments and Results}
\subsection{Setup}
The experiments that will be described will retrace the study on the model; in fact, the first tests are concerned with familiarising the parameters and functioning of the model. The first experiments were carried out with the Raabin-WBC, which enabled the best setup to be found, and this was subsequently transferred to the SC-IDB.

Two setups were considered on the Raabin-WBC: in the first setup, two classes were considered, WBC and Others, while in the second, there was only one class. In the first setup, all types of white blood cells were grouped together, including Lymphocytes, Monocytes, Neutrophils, Eosinophils and Basophils. In contrast, the Burst and Artifact types were considered in Others, whereas in the second setup, all types were grouped under the WBC class.

Both datasets include detailed annotations for segmenting each cell in the images. The SC-IDB provides two types of segmentations.

The first was performed manually: each schistocyte cell in the image is segmented by hand, outlining the cell to investigate the potential for improved results by emphasising the more pronounced angular shapes of schistocytes compared to RBCs.

The second was generated automatically, where a circumference is determined within the bounding box coordinates identifying the cell. This circumference is calculated by locating the bounding box's centre and determining the radius based on its shorter side. Using the equation of the circumcircle, the \(x\) and \(y\) coordinates of each point on the circumference are computed. This automated segmentation compares the results of manual segmentation with those of automatic segmentation, ultimately reducing the effort required for creating annotations.

\subsection{Comparative Models}
The experiments involved three OD models: the first is DE-ViT, which is this work's analysis focus, while the others, Faster R-CNN 50 and Faster R-CNN X 101, were used as comparative models. The former uses ResNet-50 as a backbone in combination with the Feature Pyramid Network (FPN), while the latter uses ResNeXt 101, again in combination with FPN, as a backbone. Both models were implemented using the Detectron2 library~\cite{detectron2}.

\subsection{Experiments}
The experiments described concern the DE-ViT model and can be divided into two macro-categories: the experiments performed on the Raabin-WBC to detect WBCs and SC-IDB to identify schistocytes. We started with the Raabin-WBC dataset to test the model's capabilities and find the best configurations. Subsequently, these configurations were used for the SC-IDB.

\subsubsection{\#WBC.1}\label{ex_1}
The first experiment was carried out following the guidelines of the reference work~\cite{ref_devit}. Thus, the prototypes of the new classes were extracted, which in this case corresponded to two classes: WBC and Others. As far as the basic and background prototypes are concerned, the default ones are used. The DE-ViT pre-trained model used for the experiment corresponds to the Vit L 14 version. In this experiment, the model was evaluated without any fine-tuning, observing the potential of the prototype mechanism.

\subsubsection{\#WBC.2}\label{ex_2}
In this experiment, the background prototypes were modified. They were extracted from the images of the training set, while other parameters remained unchanged. This allowed the model to be evaluated by adding background prototypes relevant to the working domain.

\subsubsection{\#WBC.3}\label{ex_3}
This experiment no longer uses the pre-trained model, but the Region Proposal Network (RPN) is re-trained with the images from the training set. The RPN's weights are merged with those of the Vit L 14 model to create the initial weights of the DE-ViT model. The setups of the previous experiments were retained; however, following the RPN's re-training, the previously used basic prototypes were removed.

\subsubsection{\#WBC.4}\label{ex_4}
In the last experiment conducted on the Raabin-WBC, the dataset was switched to a class: WBC. The prototypes of the WBC class were extracted, and the RPN was re-trained. Finally, the DE-ViT model was re-trained as done in the previous experiment.

\subsubsection{\#SCT.1}\label{ex_5}
Here, the best setups found in previous experiments were used. Both class prototypes were extracted using the circular mask and background prototypes. The RPN was re-trained with the current dataset, and the model was finally re-trained.

\subsubsection{\#SCT.2}\label{ex_6}
In this experiment, the previous experiment was repeated using the second type of segmentation, \ie done manually. Thus, both class and background prototypes were extracted again, and the RPN and the DE-ViT models were re-trained. 

\subsubsection{\#SCT.3}\label{ex_7}
In the last experiment, the knowledge of the model trained on the Raabin-WBC was used to identify schistocytes. The setup used comprises the trained model of the experiment \hyperref[ex_4]{\textbf{\#WBC.4}}, the basic prototypes of the Raabin-WBC dataset with one class and the prototypes of the schistocytes, with precise segmentation, as a new class. The model was not re-trained, and there was no fine-tuning. 

\subsection{Results}
\Cref{table:table_wbc,table:table_schisto} show the experiments' results on both datasets. 
\begin{table}[ht]
    \centering 
     \caption{Results of the experiments \hyperref[ex_1]{WBC.1}, \hyperref[ex_2]{WBC.2}, \hyperref[ex_3]{WBC.3}, \hyperref[ex_4]{WBC.4} conducted on the test set of Raabin-WBC. The best values are highlighted in bold.}
     \label{table:table_wbc} 
    \begin{tabular}{|l|c|c|c|c|c|c|}
    \hline
    \multicolumn{1}{|l|}{Metrics} & \multicolumn{4}{c|}{DE-ViT} & \multicolumn{1}{c|}{ FRCNN50} & \multicolumn{1}{c|}{FRCNNX101}\\
    \cline{2-5}
    \multicolumn{1}{|c|}{}& WBC.1 & WBC.2 & WBC.3 & WBC.4 & &\\
    \hline
    AP                  & 0.06  & 0.01  & 20.48 & 30.60  & 51.24           & \textbf{57.45}\\
    AP50                & 0.20   & 0.02  & 52.59 & 62.35  & 85.11          & \textbf{88.88}\\
    AP75                & 0.00     & 0.00     & 11.36 & 24.63  & 55.22           & \textbf{69.37}\\
    $AR^{maxDets=1}$    & 0.00     & 0.00     & 0.90  & 0.86   & \textbf{0.93}   & 0.60   \\
    $AR^{maxDets=10}$   & 0.00     & 0.00     & 7.13  & 7.24   & 8.22            & \textbf{8.45} \\
    $AR^{maxDets=100}$  & 1.01     & 0.30   & 36.52 & 38.77  & 57.26           & \textbf{63.74}\\
    \hline
	\end{tabular}
\end{table} 
\begin{table}[ht]
	\centering 
        \caption{Results of the experiments \hyperref[ex_5]{SCT.1} \hyperref[ex_6]{SCT.2} \hyperref[ex_7]{SCT.3} conducted on the test set of SC-IDB. The best values are highlighted in bold.} 
	\label{table:table_schisto}
	\begin{tabular}{|l|c|c|c|c|c|}
    \hline
    \multicolumn{1}{|l|}{Metrics} & \multicolumn{3}{c|}{DE-ViT} & \multicolumn{1}{c|}{ Faster R-CNN 50} & \multicolumn{1}{c|}{Faster R-CNN X 101}\\
    \cline{2-4}
    \multicolumn{1}{|c|}{}& SCT.1 & SCT.2 & SCT.3  & &\\
    \hline
    AP                  & 7.20   & 6.13  & 0.00   &  0.71 & \textbf{29.26}  \\
    AP50                & 16.84  & 14.34 & 0.00   &  2.08 & \textbf{44.37}  \\
    AP75                & 3.79   & 3.30  & 0.00   &  0.33 & \textbf{35.19}  \\
    $AR^{maxDets=1}$    & 5.54   & 5.27  & 0.00   &  1.65 & \textbf{13.93}  \\
    $AR^{maxDets=10}$   & 23.51  & 23.83 & 0.00   &  6.64 & \textbf{52.97}  \\
    $AR^{maxDets=100}$  & 44.38  & 44.82 & 0.00   &  21.31 & \textbf{57.42} \\
    \hline
	\end{tabular}
\end{table}

\subsection{Observations and Analysis}
The results obtained with DE-ViT on the Raabin-WBC, in particular, \hyperref[ex_1]{WBC.1} and \hyperref[ex_2]{WBC.2}, show that the prototype mechanism alone is not sufficient.

The main reason is the model's domain change. In fact, the weights used for the first two experiments were trained on the COCO dataset, which contains images of common objects. Such a predominant domain change may have resulted in poor performance on the part of the model.

The above is corroborated by the results obtained in the experiments \hyperref[ex_3]{WBC.3} and \hyperref[ex_4]{WBC.4}, the results of which are improved. In the latter two cases, the model was trained with images belonging to the domain, contributing to improved results.
The comparative models proved robust in the WBC identification task, obtaining better results than the DE-ViT model, even in a Few-Shot configuration.

As far as SC-IDB is concerned, the results of the experiments dropped considerably compared to the Raabin-WBC. Comparing the results of the experiments \hyperref[ex_5]{SCT.1} and \hyperref[ex_6]{SCT.2} shows that introducing a more precise segmentation did not lead to an improvement in performance. Finally, the experiment \hyperref[ex_7]{SCT.3} shows the worst results, confirming that using prototypes without model training does not lead to good results. In this group of experiments, the performance of the comparative models declined, with only the Faster R-CNN X 101 model performing well. This is probably due to the greater difficulty of the task, as the identification of schistocytes is more complex than that of leucocytes, even for domain experts.

\section{Conclusions}
This study delves into the analysis of medical images, focusing mainly on blood smears. It examines the composition of blood cells and the various diseases, disorders, and issues that affect them. Additionally, the study highlights the potential benefits of automating specific laboratory processes, which would otherwise significantly strain the expertise of laboratory technicians.

Two distinct datasets were analysed, each with specific technical characteristics and tasks: one involved the detection of leucocytes, while the other focused on identifying schistocytes.

The research then shifts to evaluating an FSOD method, the DE-ViT model. A comprehensive explanation of the model is provided, followed by a detailed account of the experiments conducted. These experiments underscore the study and reveal issues, particularly concerning domain switching.

The primary research question was whether the new FSOD method could address the challenge of having a dataset with limited labelled elements. However, the experiments indicated that this was not achievable. Unlike the leucocyte task, the model struggled with localizing schistocytes, where it performed more successfully. This discrepancy was attributed to the greater difficulty localizing schistocytes within the images. Despite the DE-ViT model's performance, the study underscores the significance of classical object detection models, which have proven highly robust and reliable, even in few-shot tasks.

Regarding future research directions, the DE-ViT model demonstrated shortcomings in handling medical image datasets. Further advancements could involve training the model with more images from the relevant domain, enabling it to better adapt to the distinctive patterns characteristic of these images.

%
%
%

\end{document}